\documentclass[superscriptaddress,showpacs,amsmath,amssymb,aps,prd,twocolumn,floatfix,nofootinbib]{revtex4-1}

\usepackage{graphicx}% Include figure files
\usepackage{dcolumn}% Align table columns on decimal point
\usepackage{bm}% bold math
\usepackage{color}
\usepackage{amsmath}
\usepackage{amssymb}
\usepackage{subfigure}
\usepackage{lineno}
\usepackage{hyperref}
\hypersetup{
    colorlinks=true,
    linkcolor=blue,
    filecolor=magenta,      
    urlcolor=blue,
}

\usepackage{todonotes}
\usepackage{hhline}
\usepackage{tabularx}
\usepackage{multirow}

\begin{document}

%\title{PILArNet: Open Sample of Simulated Liquid Argon Time Projection Chamber Data for Collaborative Software Development}

\title{PILArNet: Public Dataset for Particle Imaging Liquid Argon Detectors in High Energy Physics}

%%%%%%%%%%%%%%%%%%%%%%%%%%%%%%%%%%%%%%%%%%%%%%%%%%%%%%%%%%%%%
% List of institutions in command form:
\newcommand{\ANL}{Argonne National Laboratory, Lemont, IL, 60439, USA}
\newcommand{\SLAC}{SLAC National Accelerator Laboratory, Menlo Park, CA, 94025, USA}
\newcommand{\Tufts}{Tufts University, Medford, MA, 02155, USA}

% So that institutions appear in alphabetical order:
\affiliation{\ANL}
\affiliation{\SLAC}
\affiliation{\Tufts}

% Authors in alphabetical order
\author{C.~Adams} \affiliation{\ANL}
\author{K.~Terao} \affiliation{\SLAC}
\author{T.~Wongjirad} \affiliation{\Tufts}

\collaboration{DeepLearnPhysics collaboration}\email{contact@deeplearnphysics.org} \noaffiliation

%\linenumbers
\begin{abstract}
Rapid advancement of machine learning solutions has often coincided with the production of a test public data set.  Such datasets reduce the largest barrier to entry for tackling a problem -- procuring data -- while also providing a benchmark to compare different solutions. Furthermore, large datasets have been used to train high-performing feature finders which are then used in new approaches to problems beyond that initially defined. In order to encourage the rapid development in the analysis of data collected using liquid argon time projection chambers, a class of particle detectors used in high energy physics experiments, we have produced the PILArNet, first 2D and 3D open dataset to be used for a couple of key analysis tasks. The initial dataset presented in this paper contains 300,000 samples simulated and recorded in three different volume sizes. The dataset is stored efficiently in sparse 2D and 3D matrix format with auxiliary information about simulated particles in the volume, and is made available for public research use. In this paper we describe the dataset, tasks, and the method used to procure the sample.%We also present the forum we have setup to be able to share parts or all of an algorithm, specifically trained models. 
\end{abstract}
%\begin{abstract}
%For past problems in machine learning, the occurrence of rapid advancement often coincides with the production of a test public data set.  Such datasets reduces the largest barrier to entry for tackling a problem -- procuring data -- while also providing a benchmark for different solutions to compare with one another. Furthermore, development has also been spurred when a large dataset has been used to train high-performing feature finders that can be the basis of new approaches to problems beyond that initially defined.  In order to bring such rapid development in the problems of analyzing LArTPC images, we have produced a public dataset to be used for a couple of defined tasks. We describe the dataset and tasks and the method to procure it.  We also define the means for submitting the predictions on a test sample that will be used to benchmark between different solutions. We also present the forum we have setup to be able to share parts of all of an algorithm, specifically trained models.
%\end{abstract}

\maketitle
%\newpage
 
%\listoftodos
 
\section{Introduction}
Recently, revolutionary advancements in the field of Computer Vision (CV) have been made by machine learning (ML) techniques~\cite{AlexNet,NatureDL}. Contributing factors to this success include: advancements in machine learning algorithms and optimization methods, evolution of computing hardware such as graphic processing units, and large, public image datasets~\cite{MNIST,CIFAR,ImageNet,COCO,PASCAL,ShapeNet}. 
The datasets each consist of a large number of photographs of real world objects (i.e. a cat, dog, car, airplane, etc.), which are annotated with one of more specific image analysis tasks in mind, e.g. classification, semantic segmentation, object detection. 
This annotation provides the answers for the associated tasks in order to enable the development of {\it supervised} ML algorithms. 
One famous example is the development of deep Convolutional Neural Networks (CNNs), which grew out of the CV community's work on the large ImageNet dataset~\cite{AlexNet,vgg,googlenet,resnet,densenet,resnext}.
Accurately labeled, large statistics datasets with public availability are thus highly valuable as drivers of machine learning innovation.
% Thus, high statistics of accurately labeled image data, sometimes with single pixel-level precision, are highly valuable. 

Modern ML algorithms developed in the field of CV are being applied across many domains of science.  Examples include tomographic images of living cells~\cite{CryoEM}, the dynamics of electronic charge distribution in plasma~\cite{Plasma}, and the image formed from the spatial pattern of ionization energy deposited by charged particles traveling across a detector~\cite{UBPaper1,UBPaper2}. These images are often different from real world photographs and carry unique features associated to both the image subjects and a technology used for recording image data. As such, challenges and developments toward solution become a domain-specific effort. The dataset we describe in this paper is for data coming from a specific class of detectors used in particle physics experiments.

Liquid Argon Time Projection Chambers (LArTPCs) are a proven technology for high precision imaging of charged particles in the field of high energy physics, in particular accelerator-based neutrino oscillation experiments~\cite{ICARUS,MicroBooNE,SBN,DUNE}.  
LArTPCs are capable of recording the trajectories of charged particles in the liquid argon medium at $\approx1$~mm spatial resolution in many mega-pixel images with calorimetric information (i.e. energy deposited by a particle along the trajectory).  Despite the quality of images produced by this detector technology (and in many ways because of the complexity and detail of data), the analysis of high-resolution images of particle trajectories with the purpose of extracting high-level physics information remains challenging.  
ML techniques, in particular deep neural networks, are a promising class of solutions~\cite{UBPaper1,UBPaper2,UBNature,dunendcnn,gnnexatrk}. The published ML applications to date employ supervised learning techniques trained on large statistics samples coming from Monte Carlo simulations.
These simulations are able to produce artificial images which are increasingly better facsimiles of the real data to be analyzed.  
This is because several open source software suites~\cite{Geant4,LArSoft}, developed over decades, can provide an accurate simulation of particle interactions in matter which is crucial for providing high quality training sample with labels.

%In this paper, we present the first public open simulation dataset for LArTPCs. 
In this paper, we present the Particle Imaging in Liquid Argon dataset, PILArNet, and the first public open simulation dataset for LArTPCs as an initial contribution\footnote{\url{https://osf.io/vruzp}}.
Following the successes enabled by a large public dataset in the field of ML and CV, this effort is intended to accelerate development effort for data reconstruction and analysis techniques in the domain of particle imaging with LArTPCs.  A public dataset makes research output more transparent and seeds collaborative algorithm development across experiments and larger scientific research communities working on computer vision challenges in both academia and industries. This effort should directly benefit all particle physics experiments that employ a LArTPC detector including MicroBooNE~\cite{MicroBooNE}, ICARUS~\cite{ICARUS}, SBND~\cite{SBN}, and DUNE~\cite{DUNE}. 
DUNE, in particular, is a flagship experiment in the U.S. Department of Energy for the next decade to come, and PILArNet already enabled SBN-DUNE cross-experimental ML algorithm development~\cite{dunendcnn}.  Our contributions presented in this paper include the following:
\begin{itemize}
    \item public particle imaging data hosting tier and the first open dataset for LArTPCs;
    \item description of the first simulation dataset; and
    \item software suites for interfacing with the dataset.
\end{itemize}

%We encourage the community to join this effort and share scientific data where possible.

%In this paper, we present the first public dataset of liquid argon time projection chambers (LArTPCs), a class of imaging detector technology in the field of elementary particle physics. LArTPCs are capable of recording a trajectory of charged particles in the liquid argon (LAr) medium at $\approx1$~mm spatial resolution in many mega-pixel images with calorimetric information (i.e. energy deposited by a particle along the trajectory).  Despite the quality of images produced by this detector technology, however, analyzing detailed image of particle trajectories to extract high level physics information remains challenging.  ML techniques, in particular DNNs, are shown to be promising~\cite{UBPaper1,UBPaper2,UBNature} for bringing a solution. Following the successes enabled by a large public dataset in the CV community, we make this dataset available to accelerate data reconstruction and analysis technique development for particle physics experiments that employ a LArTPC detecotr, including MicroBooNE, ICARUS, SBND, and DUNE which is a flagship experiment in the U.S. Department of Energy for the next decade to come.  The presence of public dataset is also a key for the reproducibility of research in algorithm development.

\section{Liquid Argon Time Projection Chambers}
\begin{figure}[t]
\centering
\includegraphics[width=0.48\textwidth]{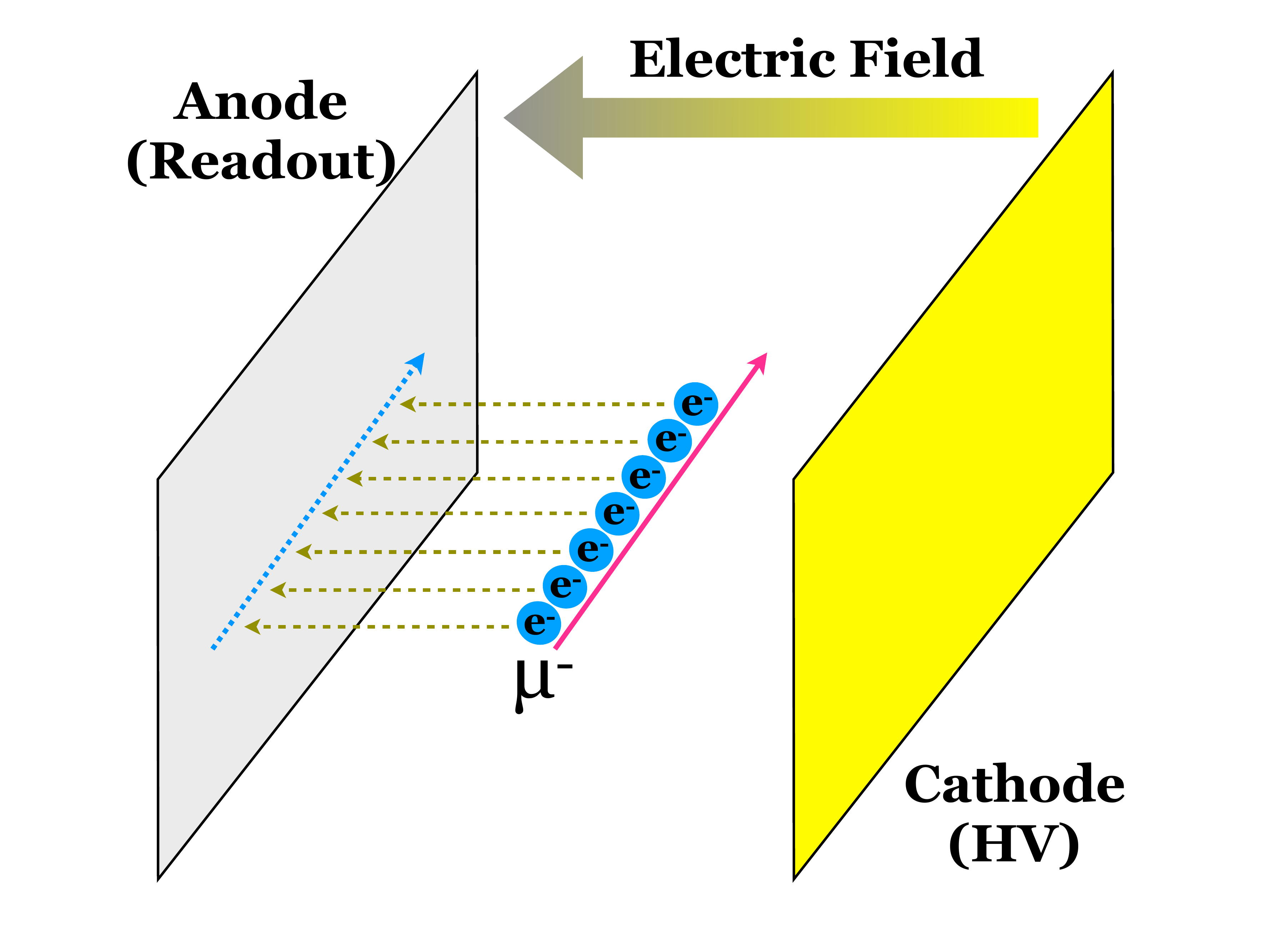}
\caption{A pictorial description of how a particle trajectory is recorded in a LArTPC, which consists of liquid argon volume sandwiched by cathode (right, yellow) and anode (left, gray) planes. When a muon traverses the argon volume, ionization electrons are created along the trajectory and subsequently drift toward the anode plane due to the electric field. The anode plane contains many electrodes for sensing the position and amount of the drifted charge. Their signals are amplified, digitized and recorded for later imaging.
\label{fig:TPCPrinciple}}
\end{figure}

In this section, we describe the central working principle of a  LArTPC detector in order to provide background as to what the images from LArTPCs capture.  
A LArTPC consists of a liquid argon volume sandwiched between anode and cathode planes. 
A large negative voltage is applied at the cathode plane in order to create a uniform electric field inside the rectangular volume. 
See Figure~\ref{fig:TPCPrinciple} for a drawing of the setup.
 %(along with and there may be additional hardware component to keep a uniform
When a charged particle traverses inside the volume (a muon in Figure~\ref{fig:TPCPrinciple}), ionization electrons are created along its trajectory. The electric field pushes the electrons toward the anode plane at a constant velocity. Critically, the anode needs to be constructed with some method to read out the amount and location of ionization electrons which have drifted to it. Detailed summaries of existing LArTPC detectors cane be found in the references~\cite{MicroBooNE,ICARUS}.% There are many more detailed summaries of the working of lartpcs, for example \todo{add some references here - uB signal processing papers?}
 %charge-sensitive readout electronics which detect the drifted ionization electrons and record data in the form of digitized waveforms. 

\subsection{Wire-based Detector}
\begin{figure}[t]
\centering
\includegraphics[width=0.48\textwidth]{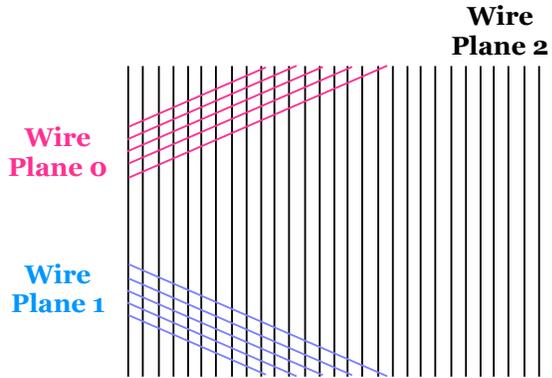}\\
\vspace{-0.2in}
\caption{An illustration of an anode for a wire-based LArTPC. In this example, the wire-based anode consists of three wire planes and only a small portion of all wires are drawn for induction planes 0 and 1. 
%\includegraphics[width=0.40\textwidth]{Anode1.pdf}
%\caption{An illustration of an anode for a wire-based (top) and pixel-based (bottom) LArTPCs. In this example, the wire-based anode consists of three wire planes and only a small portion of all wires are drawn for induction planes 0 and 1. 
\label{fig:Anode}}
\end{figure}

There are two major methods for constructing a LArTPC anode plane:  ``{\it wire}-based'' and ``{\it pixel}-based'' readouts. 
The anode for a wire-based LArTPC consists of multiple planes where each plane, called a {\it wire plane}, is made out of many parallel wires. Among the multiple wire planes, the last -- as seen from the drift electrons -- is biased with the most positive electric potential and is called the {\it collection plane}, because the drift paths of the ionization electrons 
terminate on the collection wires and cause unipolar signals. The other wire planes, which are placed in front of the collection plane, are called {\it induction planes}. Wires on each induction plane are kept at a constant voltage such that ionization electrons will pass between wires toward the collection plane. Ionization electrons, as they pass the wires, induce a bipolar current signal on the wire which is the signal to be detected. From each wire plane, the digitized waveforms from the wires are combined in order to form a 2D image of particle trajectories. The time of arrival of the charge signal gives the location of the charge deposition along the drift axis because of the approximately constant drift velocity of the ionization electrons. Wires are separated by a constant pitch distance, typically a few mm. Each wire plane has a distinct wire orientation angle (Figure~\ref{fig:Anode}) which makes each 2D image a unique projection angle.  The projection plane is the one perpendicular to the axis of the individual wires - for example, vertical wires project the 3D volume onto a plane in the horizontal axis.

The 3D shape of particle trajectories can be obtained only after combining information from multiple wire planes in the data analysis stage. Most existing and planned LArTPC detectors in neutrino oscillation experiments employ this ``wire-based'' design, including MicroBoone, ICARUS, SBND, and the DUNE far detectors~\cite{ICARUS,MicroBooNE,SBN,DUNE}. This design's wide-spread use comes from the reduced number of readout electronics per detector volume. The use of wires to record projected positions means that the number of readout channels scales roughly with the length the anode plane as opposed to its area.

\subsection{Pixel-based Detector}
The anode plane of a pixel-based LArTPC consists of a 2D array of pixel electrodes, with each channel connected to its own charge-sensitive
readout electronics. 
Example detectors of this class includes ArgonCube and the DUNE near detector~\cite{Pixel1}. The size of pixels is typically a few mm. Optimal combination of pixel size and shape is currently being investigated by R\&D programs~\cite{Pixel2}. The pixelated anode records digitized waveforms of collected ionization electrons at each pixel pad. The location of each pixel provides 2D position of drift electrons when projected on the anode plane. Given the constant velocity of drift electrons, the digitization timing can be converted into a position of drift electrons along the drift direction. In combination, a pixel-based LArTPC allows imaging of 3D particle trajectories from raw waveform without the need of 3D point reconstruction - unlike wire-based LArTPCs. This relative ease of 3D position reconstruction is one of the pixel-based readout's advantages, however the number of readout channels required scales with the area, not length, of the anode plane.

\subsection{Physics Simulation for LArTPCs}
Physics simulations of LArTPC experiments can be divided into three steps:
\begin{enumerate}
    \item  event generation, which creates a list of particles to be simulated in a detector,
    \item tracking of particles, including decay and interactions with other particles and detector medium,
    \item and simulation of the electronics signals from the result of particle tracking.
\end{enumerate}
For most  LArTPC experiments, the simulation routines are implemented within {\sc LArSoft}~\cite{LArSoft}, a community-driven software framework for simulation and data analysis. While {\sc LArSoft} is open-source and many of the components are shared across experiments, each experiment must still specify many details such as detector materials, readout configurations, and more in order to build their respective simulation. 
Consequently, the experiment-specific implementations of LArSoft are not typically open-source. The experiment-specific components manifest primarily in step 1 and 3 in the list above, because each experiment has a different physics signal to look at as well as different detector configurations. On the other hand, step 2 above is almost identical among experiments except for the fact that the geometry of liquid argon volume may be different. This step uses {\sc Geant4}~\cite{Geant4}, a software developed over many decades in the field of High Energy Physics for tracking particle interactions in the detector medium (i.e. liquid argon). 

In this paper, we present a data sample generated using our custom event generator (step 1) followed by a {\sc Geant4} particle tracking simulation run within {\sc LArSoft} (step 2). This release of sample does not incorporate step 3 for which we need a dedicated effort to develop open-source algorithms for a hypothetical detector response. Instead, we applied spatial smearing of charge to every voxel in order to mimic a real detector resolution. This will be described in the following sections. While it lacks accurate detector-specific effects on signal, the generated images capture the geometrical and basic calorimetric features of particle trajectories in a LArTPC detector. The sample can be used for experiment-agnostic algorithm development including ML applications for reconstruction and analysis of LArTPC data.

\section{2D/3D LArTPC Simulation Samples}
\begin{figure*}[t]
\centering
\includegraphics[width=0.48\textwidth]{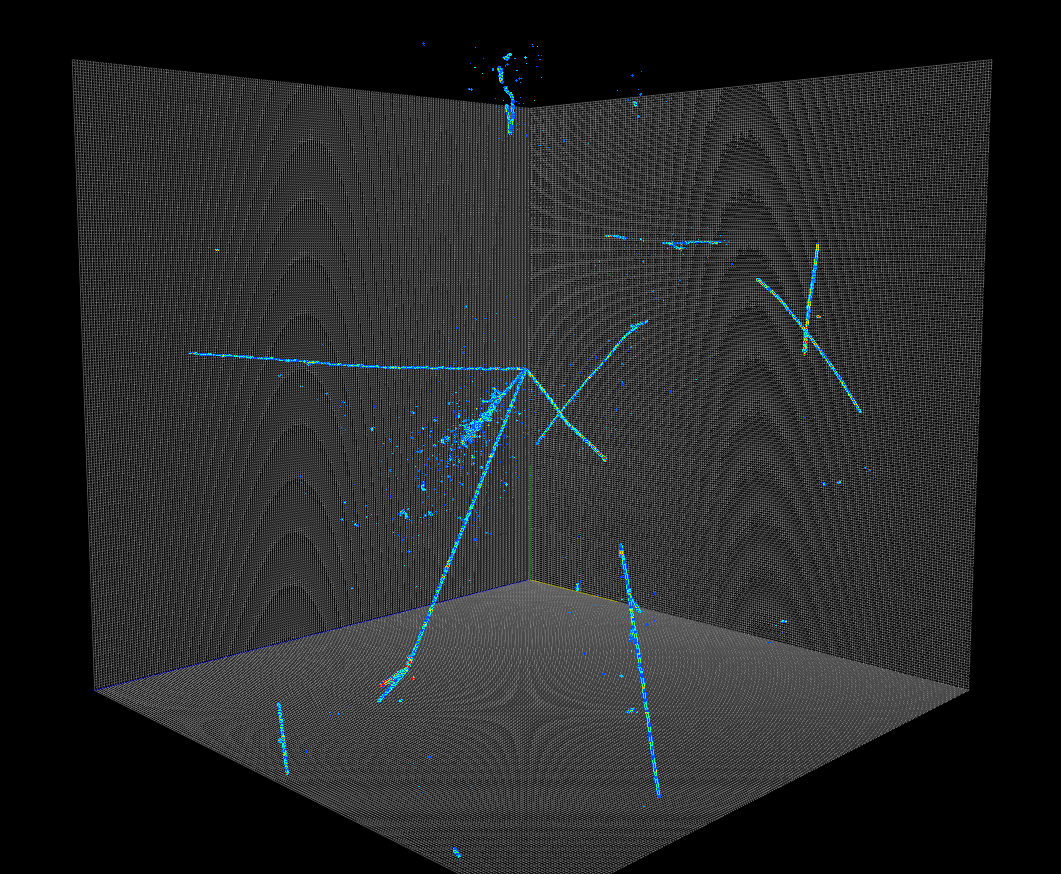}
\includegraphics[width=0.47\textwidth]{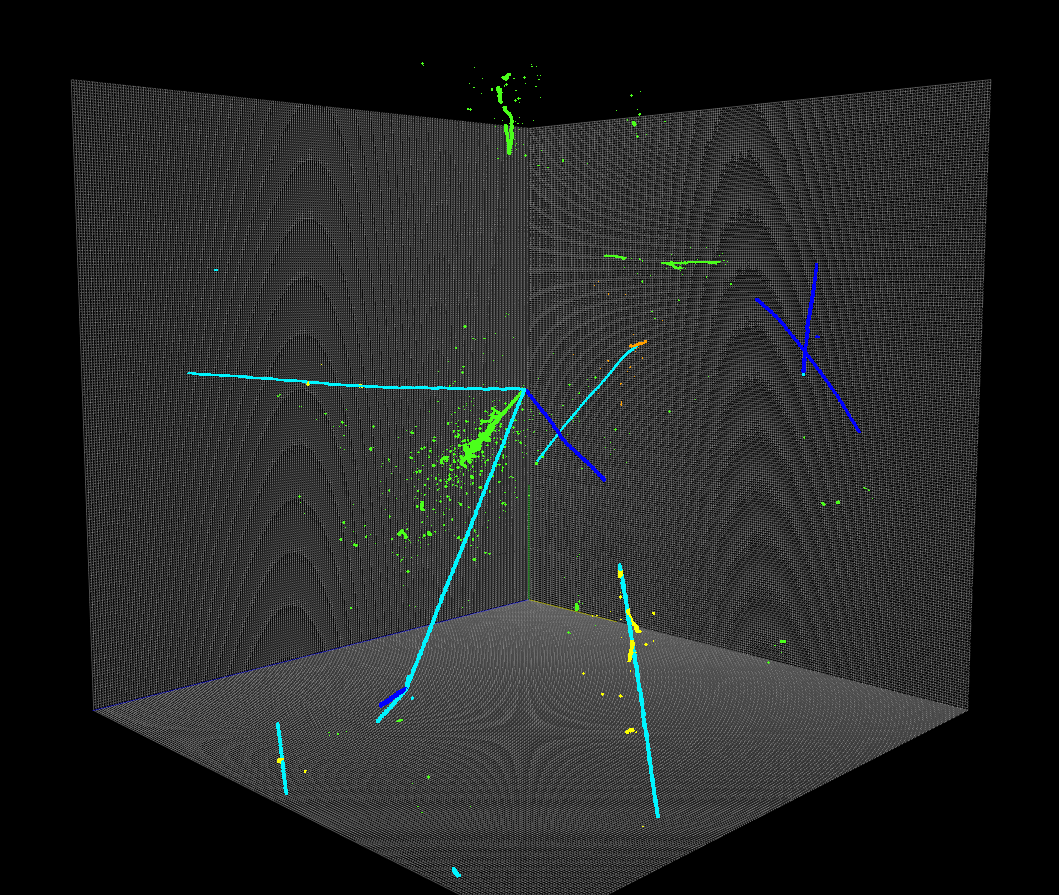}
\caption{Recorded energy deposition at the end of particle tracking simulation, visualized using the {\sc LArCV} native data visualization tool. Left: energy deposition recorded per cubical cells. The color scale changes from dark blue,  cyan, yellow, and red for increasing amount of energy per cell. Right: cells categorized into 5 colors based on the most dominating particle type in terms of deposited energy per cell. Five categories are protons (dark blue), Michel electrons (orange), delta rays (yellow), electromagnetic showers (green), and other particles (cyan) typically minimum ionizing particles such as muons and pions. 
\label{fig:demo}}
\end{figure*}
Our dataset is a simulation of particles created for the development of data analysis techniques for LArTPC detectors, and contains a total of 300,000 physics events generated using custom event generators~\footnote{\url{https://github.com/DeepLearnPhysics/LArTPCEventGenerator}} followed by a particle tracking simulation using {\sc Geant4}. The samples are all independently simulated events, and consist of three different cubical volumes including $192^3$, $512^3$, and $768^3$ pixels, 100,000 samples each. In each event, two event generators, referred to as {\sc MultiPartVertex} (MPV) and {\sc MultiPartRain} (MPR), are used to generate a list of particles for tracking simulation. In this section, we describe the simulation process, recorded information, and data access methods.

%% Taritree got here for now

\subsection{Event Generator Simulation}
In each event, the MPV generator simulates N particles all originating from a unique 3D point, called the {\it vertex}, where N is randomly set to an integer value between 1 and 6 with equal probability. There are five categories of particles that can be generated: electron, photon, (anti-)muon, charged (anti-)pion, and proton. Each category has two simulation parameters for generation: the range of kinetic energy and the maximum number of multiplicity per category as shown in Table~\ref{tab:MPV}. N particles are randomly drawn from available categories where multiplicity limit is not reached. The kinetic energy of a particle is determined within the specified range, and is distributed uniformly. Among the five categories, muon and pion can be either a particle or an anti-particle with equal probability. The reason for including both types is because the behaviors are different: stopped anti-muons decay 100\% of the time with a positron emmission while stopped muons get captured by an argon nucleus more than half the time in LArTPCs. Thus including both types is possibly of interest for data reconstruction algorithm development. Neither existing nor future LArTPCs plan to observe anti-protons.
\begin{table}[t]
    \centering
    \begin{tabular}{|c|c|c|}
    \hline
    Category & Max. Multiplicity & Kinetic Energy [MeV] \\
    \hline
    Electron & 1 & 50 to 1000 \\
    \hline
    Photon   & 3 & 50 to 1000 \\
    \hline
    (Anti-)Muon & 1 & 50 to 1000 \\
    \hline
    (Anti-)Pion & 2 & 50 to 1000 \\
    \hline
    Proton   & 2 & 50 to 400 \\
    \hline
    \end{tabular}
    \caption{MPV generator configuration parameters including the maximum particle multiplicity limit per category and the range of kinetic energy allowed. Within this range the kinetic energy value is randomly assigned.}
    \label{tab:MPV}
\end{table}

\begin{table}[t]
    \centering
    \begin{tabular}{|c|c|c|}
    \hline
    Category & Max. Multiplicity & Kinetic Energy [MeV] \\
    \hline
    Electron & 5 & 50 to 1000 \\
    \hline
    (Anti-)Muon & 5 & 50 to 1000 \\
    \hline
    Proton   & 5 & 50 to 400 \\
    \hline
    \end{tabular}
    \caption{MPR generator configuration parameters  including the maximum particle multiplicity limit per category and the range of kinetic energy allowed. Within this range the kinetic energy value is randomly assigned.}
    \label{tab:MPR}
\end{table}
The MPR generator is similar to MPV except generated particles are not required to come out of the same vertex. The total number of particles generated by MPR is fixed to 10 in every event, and the start point is uniformly distributed in the simulated volume of liquid argon. Table~\ref{tab:MPR} describes the list of categories and their configurations for MPR generator used for the public sample. The categories include electron, (anti-)muon, and proton. For MPR, the probability of each category to be chosen for a particle generation is not uniform across categories. It is set to 20\%, 60\%, and 20\% for electron, (anti-)muon, and proton respectively. There is a weak motivation to populate the space with (anti-)muons which might make the simulated sample look closer to surface LArTPC detectors that are exposed to cosmic ray particles. %\todo{Why not pions and photons in this sample?}

\subsection{Particle Tracking Simulation}
Following the generation of particles, tracking simulation is done using {\sc Geant4}, version {\sc v4.10.1.p03} with physics list {\sc QGSP\_BIC}. within {\sc LArSoft}. {\sc Geant4} propagates all particles and simulates both energy depositions in the medium as well as secondary interactions induced by the generated particles. In this simulation stage, the liquid argon volume is {\it voxelized} into many cubes of the same size with 3~mm side. Ionization energy deposition by charged particles traversing cells of the voxelized volume is recorded as a sparse matrix data for individual particles. This allows us to track energy depositions per cell per particle.  In addition to energy deposition information, we also record summary information per particle that is generated by either an event generator or through secondary particle interactions during the tracking stage.

The recorded volume of the tracking simulation may be one of three different sizes: $192^3$, $512^3$, and $768^3$ pixels. In each case,  the spatial resolution of each voxel is 3~mm. The location of the recorded cubic volume is determined under two conditions. First, it must contain the vertex defined by MPV event generator. Second, it must maximize the energy depositions of particles generated by MPV. Under these conditions, the position of the volume is set randomly. By construction, therefore, the MPV vertex can be found in an each event. 

\subsection{Spatial Smearing of Energy Deposition}
The output of the particle tracking simulation is a collection of summary information per particle, and energy depositions per particle per cell of a voxelized volume.  We refer to those outputs {\it particle} and {\it particle-cluster} information. For each pixel, smearing of the energy deposition is applied using a normal distribution with a spread of a unit voxel (3~mm). This makes a particle trajectory thicker as energy from a pixel is spread to its neighbor pixels. For every pixel, the value is computed again by summing contributing energy from neighbor pixels as a result of smearing. The sum of all pixel values remains the same before and after the smearing process.

\subsection{Image of Particle Trajectories}
 From the smeared energy depositions, we generate 3D energy deposition information, referred to as {\it energy-3D information}, in the detector by summing energy depositions per cell across all contributing particles.  This is shown in the left of Figure~\ref{fig:demo}, and resembles what a pixel LArTPC can detect in an ideal world where detector responses including readout electronics, electric field response, and recombination of ionization electrons are absent. While diffusion of ionization electrons in drift is not simulated, an artificial smearing of deposited energy is applied. Despite the lack of those simulation stages, topological and geometrical features in 3D images that pose challenges to LArTPC data reconstruction are still present. 

In addition to energy-3D, {\it segment-3D}, which holds a category of particle types per pixel, is also created from particle information. 

The segment-3D image is only different from the energy-3D by individual pixel values. The pixel values vary between 0 to 4 in integer steps and defined as follows.
\begin{enumerate}
	\item[0:] Protons (a heavily ionizing particle), which have a ``track'' topology (1D line-shaped trajectory).
	\item[1:] Other particles (typically minimum ionizing particles) with a ``track'' topology: (anti-)muon and charged (anti-)pion.
	\item[2:] Electromagnetic (EM) particles (i.e. electrons, positrons and photons) which have a ``shower'' topology (containing many ``Y''-shaped branches) . 
	\item[3:] Michel electrons, a decay electron and positron from a muon and anti-muon respectively.
	\item[4:] Delta rays, knock-off electrons from track particles traversing the LAr volume.
\end{enumerate}
The segment-3D information can be used to train an algorithm for pixel-level categorization, a task referred to as semantic segmentation in the field of CV, and is crucially useful for LArTPC data analysis~\cite{UBPaper2}.

Finally, the 2D projections of energy-3D and segment-3D are created, referred to as energy-2D and segment-2D in this paper. Those projections are simple 2D projections along the principle 3D coordinate axis including x-y, y-z, and z-x plane projections. The contents of the output files can be used to train ML algorithms for all of tasks in the list of bullets above.

\subsection{Simulation Output}

\begin{figure}[ht]
\centering
\includegraphics[width=0.48\textwidth]{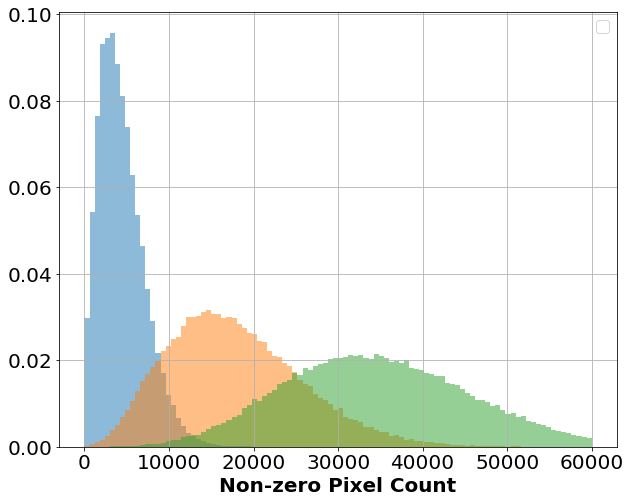}
\includegraphics[width=0.48\textwidth]{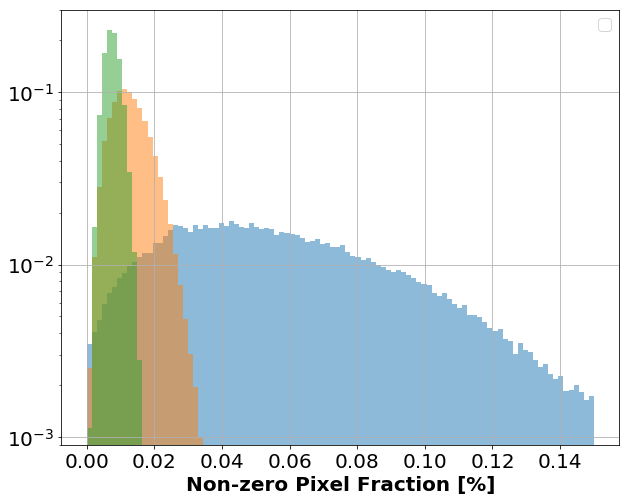}
\caption{LArTPC image data is generally sparse although locally dense. The top figure shows the number of total pixel count per 3D image. The blue, orange, and green colors of histograms correspond to $192^3$, $512^2$, and $768^3$ image size. The bottom figure shows the fraction of non-zero pixels in the cubic volume.}
\label{fig:sparsity}
\end{figure}
Sparse 2D and 3D matrix data representations in the {\sc LArCV}\footnote{\url{https://github.com/DeepLearnPhysics/larcv2}} software framework are used for efficient storage of all matrix data which includes particle-cluster, energy-3D, segment-label-3D, energy-2D, and segment-label-2D. LArTPC image data is generally sparse, but locally dense (i.e. no gap in a recorded particle trajectory). Figure~\ref{fig:sparsity} shows that, for energy-3D, the fraction of non-zero pixels in three simulated volume sizes are well below 1~\%. Moreover, because in most cases a particle trajectory is locally a one-dimensional line, the increase in the number of non-zero pixels does not scale in the same manner as the total number pixels required for a volume to contain a longer trajectory. Therefore a zero-suppressed matrix data representation is crucial to keep a reasonable storage data file size.

\subsection{Data Challenges}
\begin{table*}[t]
    \centering
    \bgroup
    \def\arraystretch{1.3}
    \begin{tabular}{|l|l|l|l|}
    \hline
        Physics Task & Input Data & Label & Computer Vision Task \\
        \hline
        Particle ID & particle-cluster & particle & Image-level classification (multinomial logistic regression) \\
        \hline 
        Particle Kinematics & particle-cluster & \multirow{2}{*}{particle} & \multirow{2}{*}{Image-level regression} \\ 
        (energy, direction) & or energy-2D/3D  & & \\
        \hline 
        Particle Trajectory & \multirow{2}{*}{energy-2D/3D} & \multirow{2}{*}{particle-cluster} & \multirow{2}{*}{Object instance segmentation (pixel clustering)}\\
        Reconstruction & & & \\
        \hline
        Position/Vertex & \multirow{2}{*}{energy-2D/3D} & \multirow{2}{*}{particle} & \multirow{2}{*}{Object detection / position regression} \\
        Reconstruction & & & \\
        \hline 
        3D Point & \multirow{2}{*}{energy-2D} & \multirow{2}{*}{energy-3D} & \multirow{2}{*}{2D to 3D image reconstruction (inverse process)}\\
        Reconstruction & & & \\
        \hline 
    \end{tabular}
    \egroup
    \caption{Physics tasks with input data and corresponding tasks for CV techniques development. The label column shows the label information that can be used for supervised training method. }
    \label{tab:challenges}
\end{table*}
The challenge of topological LArTPC data reconstruction is to start from energy-2D or energy-3D information and derive higher level information including:
\begin{itemize}
    \item a pixel-level particle category information (i.e. segment-3D data),
    \item a cluster of pixels corresponding to individual particle trajectory (i.e. particle-cluster data),
    \item particle-level information including a particle type, momentum, start/end points and directions (i.e. particle data),
    \item image-level information including total energy, the interaction vertex, and hierarchical correlation (i.e. ``flow'') of particles that can be retrieved from particle data. 
\end{itemize}
These tasks follow the order of a typical data analysis workflow for LArTPC images, and each has an analog to an image analysis task in CV. Our dataset provides necessary information (i.e. ``labels'') for supervised training of ML algorithms. Table~\ref{tab:challenges} summarizes a set of challenges associated with this dataset, necessary label information from the dataset for supervised learning, and corresponding tasks in CV.

%In addition, we prepared {\it segment-label-3D} information for a semantic segmentation task, which has been explored extensively in LArTPC community recently~\cite{UBPaper2}. In this task, five categories are defined including proton, Michel electron, hard-scatter atomic electrons (delta-rays), electromagnetic showers, and other non-EM particles including muons and pions. 

\subsection{Hosting and Organization of Data}
Our dataset is hosted currently by the Open Science Framework (OSF) under {\it CC-by-Attribution 4.0 international} license. The OSF is one of several free data hosting tiers that can be directly cited for publications in major journals.  It also includes features to generate the digital object identifier (DOI) per data set. In addition, it provides a rich set of web browser and command-line interfaces for simple upload and download of data files.

\begin{table}[t]
    \centering
    \begin{tabular}{|c|c|}
    \hline
        Sub-folder Name & Contents \\
        \hline
        {\sc cluster}  & particle-cluster\\
        \hline
        {\sc data}     & energy-3D, segment-3D\\
        \hline
        {\sc data-2d}  & energy-2D, segment-2D\\
        \hline
        {\sc particle} & particle\\
        \hline
    \end{tabular}
    \caption{Sub-folder name under {\sc 192px}, {\sc 512px}, and {\sc 768px} top folders of the dataset and the contents in files.}
    \label{tab:foldername}
\end{table}

%\todo{This section needs to be tied back to the physics needs of lartpcs.  So, challenges listed under the Simulation Output section need to be shown in the dataset as something for which training data exists.  In particular, I think this calls for pictures and visualization.  I will work on generating images.}

The OSF data storage is organized in a hierarchical manner. The top level storage of the dataset contains three folders named as {\sc 192px}, {\sc 512px}, and {\sc768px} which contain datasets with corresponding volume size following the folder name. There are four sub-folders under each of them: {\sc cluster}, {\sc data}, {\sc data-2d}, and {\sc particle}. These sub-folders hold data files where the contents of each data file are summarized in Table~\ref{tab:foldername}. Under each sub-folder, there exist same number of data files with a common file suffix ``XX.root'' where XX is a zero-filled two digit integers to match the contents of files across sub-folders. There are 10 files under each sub-folder of {\sc 192px} and {\sc 512px}, and each file contains 10,000 events, which makes 100,000 events from 10 files for those two volume sizes. Under the sub-folders of {\sc 768px}, there are 20 files where each file contains 5,000 events. The reason for splitting into more files for this volume size is the maximum single file size limitation imposed by OSF, which is 5~GB/file. 

\section{Interacting with the Dataset}
Accessing the contents of the dataset can be done using {\sc LArCV}, which is primarily written in {\sc C++} with an extensive {\sc Python} API. {\sc LArCV} is maintained by the Deep Learn Physics (DLP) organization and software containers are distributed for algorithm development and data analysis use. The currently supported container types are {\sc Docker}~\cite{Docker} and {\sc Singularity}~\cite{Singularity}. A dedicated data interface API is also prepared for this open data set, and is described in this section.

\subsection{Software Containers}
Container images can be found in the DLP repositories hosted by {\sc docker-hub}\footnote{\url{https://cloud.docker.com/u/deeplearnphysics/repository/docker/deeplearnphysics/larcv2}} and {\sc singularity-hub}\footnote{\url{https://www.singularity-hub.org/collections/459}}. The recipe files used to build these containers can be found in DLP software repositories {\sc larcv2-docker}\footnote{\url{https://github.com/DeepLearnPhysics/larcv2-docker}}
 and {\sc larcv2-singularity}\footnote{\url{https://github.com/DeepLearnPhysics/larcv2-singularity}} respectively. Built image tags and versions are closely mirrored between images among those two container types. %For the clarity, we focus our discussion only on {\sc Docker} images in the following. 

%\begin{table}[t]
%    \centering
%    \begin{tabular}{|c|c|c|}
%        \hline
%        Tag Name & Size [GB] & ML Libraries\\
%        \hline
%        minimal & 0.36 & N/A \\
%        \hline
%        cuda90-tf1.12.0 & 3 & Tensorflow, Scikit-learn\\
%        \hline
%        cuda90-torch0.4.1 & 3 & Pytorch, Scikit-learn\\
%        \hline
%    \end{tabular}
%    \caption{Docker image tag names, their sizes, and included ML libraries.}
%    \label{tab:libs}
%\end{table}
%Currently there are three recommended {\sc Docker} image tags to be used for interfacing PubLAr datasets: {\it minimal}, {\it cuda90-tf1.12.0}, and {\it cuda90-torch0.4.1}. The size and contained ML libraries for each tag are summarized in Table~\ref{tab:libs}. In addition to ML libraries, the latter two tags incorporate a large set of scientific {\sc Python} libraries including numpy, scipy, pandas, matplotlib, pyqtgraph, plotly, and more. The detailed list can be found in the {\sc Docker} image recipe files cited in the reference. All of these containers \todo{Finish this thought}

\subsection{Data Interface APIs}
\begin{figure*}[t]
\centering
\includegraphics[width=0.95\textwidth]{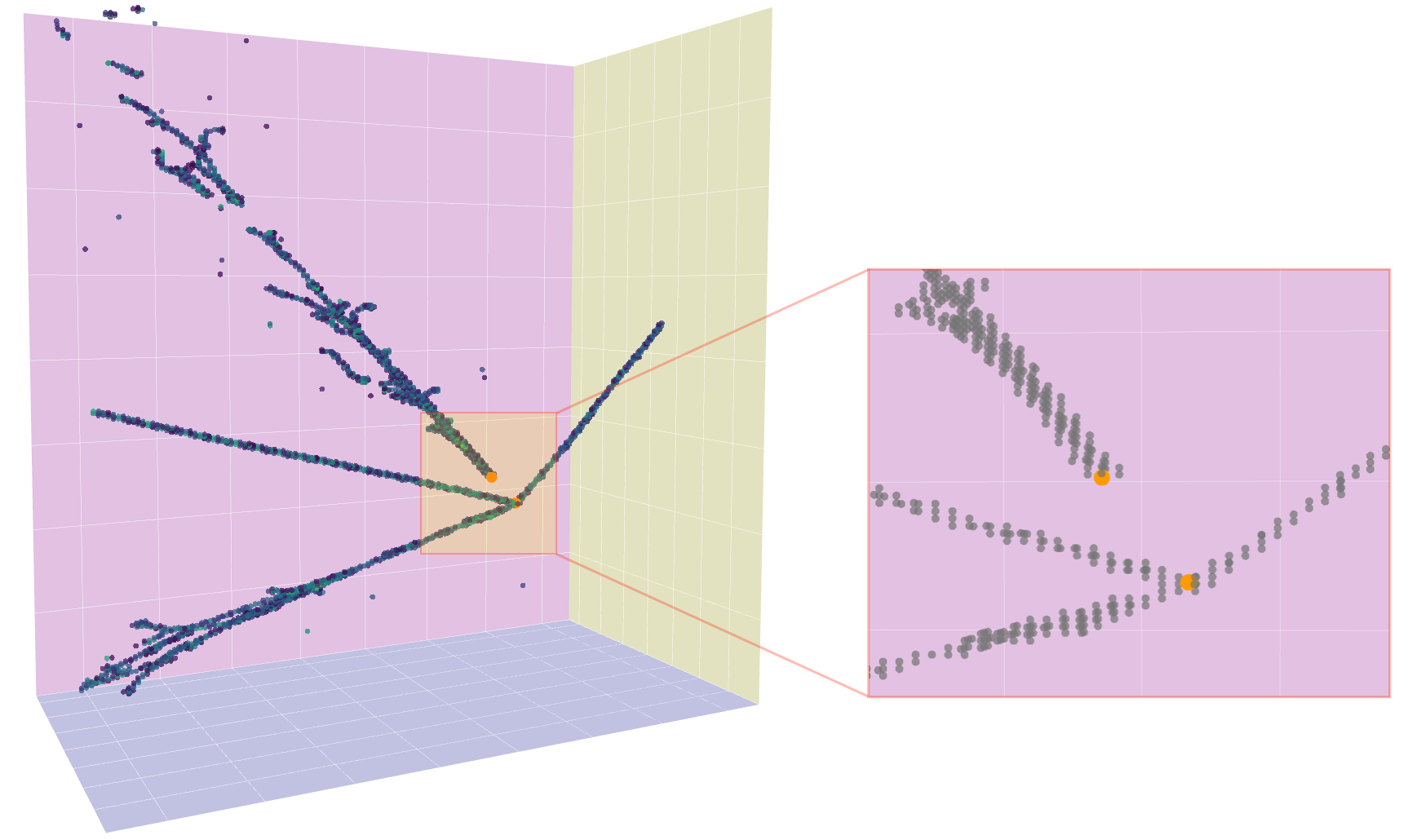}
\caption{Visualization of a particle trajectories (energy depositions) from a dataset sample using prepared {\sc Python} APIs. The orange sphere indicates the start points of particles, which can be used for supervised training of ML algorithms to regress the three-dimensional start position of particles.
\label{fig:apidemo}}
\end{figure*}

There are two recommended ways to open and interpret data files under the software container environment. A straightforward option is to use {\sc Python} API functions from the {\sc dlp\_opendata\_api} repository\footnote{\url{https://github.com/DeepLearnPhysics/dlp_opendata_api}}. The example script included in the repository demonstrates how one can open a file, move a data pointer among different samples, and retrieve data in numpy array data format, which is one of the most popular representations with connections to ML libraries. The example also shows how to match 3D imaging data with particle meta information, such as start and end points of a trajectory using plotly data visualization tool. Figure~\ref{fig:apidemo} shows an example visualization of a sample using plotly and numpy array of particle energy depositions and start positions obtained using the APIs. A similar visualization example is included in the prepared demo script.

Separately, {\sc LArCV} provides methods to interface to data using both {\sc Python} and {\sc C++}. While there is more overhead cost in learning the {\sc LArCV} framework, this method provides multi-threaded routines for reading data from files and loading to computer memory. This scheme supports configurable options to load data in either dense or sparse matrix format. While these routines are written in {\sc C++}, APIs are provided also in {\sc Python}. Fast data loading using native APIs in the {\sc LArCV} framework may be useful for training ML algorithms and running fast inference on a large set of data samples.  A python interface is also provided to serve as a distributed IO interface for MPI based applications such as distributed learning. Examples on how to use {\sc LArCV} file reader APIs for training deep neural networks is described in a tutorials\footnote{\url{http://deeplearnphysics.org/Blog/tutorial\_summary.html}} and wiki\footnote{\url{https://github.com/DeepLearnPhysics/larcv2/wiki}}.

\section{Conclusion}
We have presented PILArNet and the first open LArTPC dataset of simulated particle trajectories as an initial contribution. The first set presented in this paper contains 300,000 sample statistics with various labels that can be used for supervised ML algorithm training. The dataset is produced using {\sc Geant4} and {\sc LArSoft}, common public software frameworks in the community of LArTPC experiments, and stored in {\sc LArCV} files with a sparse matrix data representation which serves well for sparse particle energy depositions in LArTPC detectors. The dataset is hosted within the Open Science Framework, which is a recognized scientific data sharing tier highly integrated with industrial cloud services. The software containers are the modern approach to establish an identical software execution and development environment, and therefore they improve reproducibility of research. Our Docker and Singularity software container images are made available to ease the preparation of a data analysis environment.  

Recent advancements in ML have a potential to address many challenges in science research. The open datasets have been the core of evolution in the field of ML, and are most effective ways to form a connection between domain experts and researchers from outside fields. PILArNet follows this model, and it is aimed to seed cross-experimental, cross-disciplinary collaborations in order to address challenges in analyzing data of LArTPC and other particle imaging detectors. PILArNet will continue to evolve in both variety and maturity to aid domain-specific ML technique adaptations and to enable physics discoveries.

%%%%%%%%%%%%%%%%%%%%%%%%%%%%%%%%%%%%%%%%%%%%%%%%%%%%%%%%%%%%%%%%%%%%%%
%
% Acknowledgement
%
%%%%%%%%%%%%%%%%%%%%%%%%%%%%%%%%%%%%%%%%%%%%%%%%%%%%%%%%%%%%%%%%%%%%%%

\section{Acknowledgement}
This work is supported by the U.S. Department of Energy, Office of Science, Office of High Energy Physics, and Early Career Research Program under Contract DE-AC02-76SF00515. This research used resources of the Argonne Leadership Computing Facility, which is a DOE Office of Science User Facility supported under Contract DE-AC02-06CH11357.

%%%%%%%%%%%%%%%%%%%%%%%%%%%%%%%%%%%%%%%%%%%%%%%%%%%%%%%%%%%%%%%%%%%%%%
%
% Bibliography
%
%%%%%%%%%%%%%%%%%%%%%%%%%%%%%%%%%%%%%%%%%%%%%%%%%%%%%%%%%%%%%%%%%%%%%%
\newpage 
\bibliography{reference}

\end{document}